\documentclass[10pt,journal,twoside]{IEEEtran}

\usepackage{amsmath}
\usepackage{cite}
\usepackage{graphicx}
\usepackage{amsfonts}
\usepackage{latexsym}
\usepackage{subfigure}
\usepackage{algorithm}
\usepackage{algorithmic}
\usepackage{color}
\usepackage{times}
\usepackage{epsfig, graphics}
\usepackage{bm}
\usepackage{subfigure}
\usepackage{graphicx,wrapfig,lipsum}
\usepackage{amsthm}

\begin{document}

\newtheorem{theorem}{Theorem}
\newtheorem{lemma}{Lemma}
\newtheorem{conjecture}{Conjecture}
\newtheorem{corollary}{Corollary}
\newtheorem{definition}{Definition}
\newtheorem{scheme}{Scheme}
\newcommand{\argmax}{\arg\!\max}
\newcommand{\pound}{\operatornamewithlimits{\gtrless}}
\IEEEoverridecommandlockouts

\title{Coherent Communications in Self-Organizing Networks with Distributed Beamforming
\vspace{8mm}}
\author{Yi Shi and Yalin E. Sagduyu
\thanks{Yi Shi is with Virginia Tech, Blacksburg, VA 24061, USA, and Intelligent Automation, Inc., Rockville, MD 20855, USA (e-mail: yshi@vt.edu). Yalin Sagduyu is with Intelligent Automation, Inc., Rockville, MD 20855, USA (e-mail: ysagduyu@i-a-i.com).}
\thanks{This research was supported under contract D17PC00022, from Defense Advanced Research Projects Agency (DARPA).}
\thanks{Distribution Statement A (Approved for Public Release, Distribution Unlimited). The views, opinions and/or findings expressed are those of the author and should not be interpreted as representing the official views or policies of the Department of Defense or the U.S. Government.}
\thanks{\textsuperscript{\copyright}2019 IEEE. Personal use of this material is permitted. Permission from IEEE must be obtained for all other uses, in any current or future media, including reprinting/republishing this material for advertising or promotional purposes, creating new collective works, for resale or redistribution to servers or lists, or reuse of any copyrighted component of this work in other works.}
}

\maketitle

\begin{abstract}
Coherent communications aim to support higher data rates and extended connectivity at lower power consumption compared with traditional point-to-point transmissions. The typical setting of coherent communication schemes is based on a single data stream with multiple transmitters and a single receiver. This paper studies the case of multiple concurrent data streams, each with multiple transmitters and multiple receivers, in self-organizing wireless networks. A distributed optimization solution based on joint network formation and beamforming is developed for coherent group communications in a network of nodes that need to communicate over long distances. This solution significantly improves the power gain for a single data stream and the signal-to-interference-ratio for multiple data streams compared to point-to-point communications. These gains are in turn translated to improvements in communication range, power efficiency, reliability, and throughput. In this multi-layer network optimization solution, nodes coordinate with each other in a distributed manner to form transmitter and receiver groups, and communicate with each other coherently over long distances. The coherent beamforming protocol is optimized for given transmitter and receiver groups, whereas the network formation protocol is optimized to determine these groups. By using single-antenna communication nodes, the proposed optimization solution provides a major gain in  network communications and outperforms other benchmark combinations of beamforming and network formation protocols.
\end{abstract}



\section{Introduction}


Wireless communications serve various mobile applications such as multimedia, remote sensing, and inter-vehicle communications that demand high data rates over long distances but they often need to run on battery-operated devices with limited transmit powers. Many advanced techniques such as directional and multiple-input-multiple-output (MIMO) communications  have been developed to support high-rate communications with low power consumption over long distances,
which cannot be met by traditional point-to-point transmissions (from one transmitter to one receiver).
Among these techniques, \emph{coherent communications} use only a single omnidirectional antenna per node and has emerged as a viable solution to improve the \emph{power gain} between distant groups of nodes, which consequently leads to the better signal quality at receivers \cite{Bidigare13,Mudumbai09,Mudumbai10,Ochiai05}.
This power gain is up to $N^2 M$ for $N$ transmitters communicating with $M$ receivers when the number of data streams $K$ is $1$.
Examples for coherent communications include inter-vehicle communications, air-to-ground and ground-to-air communications (such as involving UAVs). One particular application that benefits from maximizing the received power is communications under jamming where high SINR needs to be maintained to balance destructive jamming effects.

A typical assumption on coherent communications is that there is one data stream and there is one receiver for this data stream, i.e., $K=1$ and $M=1$. In this paper, we consider the extension to multiple  data streams ($K>1$) operating simultaneously and potentially interfering with each other. There are multiple transmitters ($N>1$) and multiple receivers ($M>1$). For each data stream, a source node selects some of these transmitters to communicate to some of the receivers selected by the corresponding destination. This extension for self-organizing wireless networks requires joint optimization of beamforming and network formation in a distributed manner.

In a wireless network where nodes are not physically connected, it is necessary to coordinate nodes with each other to form transmitter and receiver groups for \emph{distributed coherent communications}.
Without using any centralized controller, coherent communication is supported with three protocol steps.
First, nodes self-organize themselves into transmitter and receiver groups in an ad hoc network setting without a centralized controller, where the source nodes is in the transmitter group and the destination node is in the receiver group.
Second, nodes exchange data and channel condition in the transmitter group and transmit with coherent beamforming to the receiver group.
Third, nodes receive and exchange signals in the receiver group to decode data at the destination.
We define a time frame for coherent beamforming with the above steps, which includes the data exchange in the transmitter group, inter-group communications, and the data exchange in the receiver group.

To enable coherent group communications, we develop a distributed optimization solution that combines a \emph{distributed transmit beamforming protocol} for given transmitters and receivers at the PHY layer (that sets the initial phase value at each transmitter to achieve phase coherence at receivers) and a \emph{distributed network formation protocol} at the network layer (that selects and maintains a transmitter group for each source and a receiver group for each destination).
Our approach is first building a basis in theoretical system model analysis on power gain, then designing a distributed coherent beamforming protocol (given a transmitter group and a receiver group), and finally developing joint beamforming and network formation protocols through a \emph{multi-layer network optimization} solution across PHY and network layers.
\begin{itemize}
\item \emph{Distributed coherent beamforming protocol.}
We develop a distributed coherent beamforming protocol for given a group of $N$ transmitters and a group of $M$ receivers.
The coherent communication gain is defined as the ratio of the total received energy by coherent communications and the received energy by point-to-point transmissions.
Existing coherent beamforming protocols are mainly designed for $M=1$ (single receiver) and the optimal solution with the maximum gain can be achieved if phase coherence is achieved at the receiver.
For the general case of $M>1$, it is not clear which phase coherent solution can achieve the maximum coherent communication gain.
To solve this problem, we first derive a closed-form expression of the coherent communication gain for any solution and then analyze its upper bound and assess the conditions to achieve this upper bound. We construct an optimization problem to maximize the coherent communication gain.
We then present several coherent beamforming protocols and compare their performance in simulation. We show that \emph{iterative optimization} protocol achieves the best gain that is very close to the theoretical upper bound.

\item \emph{Analysis and simulation on coherent beamforming.}
We analyze the synchronization requirement for coherent beamforming and show that only transmitters should be synchronized.
We analyze the impact of Doppler spread and show the tolerable Doppler spread as a function of distance between transmitter and receiver groups.
Then we analyze the effects of group size, inter-group distance, and channel model for the developed coherent beamforming protocols.
We show that the iterative optimization protocol achieves the best performance among the developed protocols and achieves near-optimal performance by comparing with the upper bound.
Then we study the performance over distance and find the advantage of coherent beamforming as the inter-group distance increases.

\item \emph{Distributed network formation protocol.}
We develop a protocol at the network layer that selects transmitters and receivers for each source-destination pair to improve the coherent communication gain.
Existing designs for coherent communications mainly focus on the case of a single source-destination pair ($K>1$).
Since there is only one source and one destination, all nodes close to the source operate as transmitters and all nodes close to the destination operate as receivers.
With multiple source-destination pairs, nodes need to be assigned to sources/destinations.
Moreover, there is interference from data streams for other sources.
We use the signal-to-interference-ratio (SIR) as the performance measure and define an objective function based on multiple SIRs for all destinations.
While the exhaustive search on network formation can find the optimal solution that maximizes this gain, this solution can only be used as a benchmark due to its high complexity.
Therefore, we design and compare low-complexity and distributed protocols for network formation and beamforming. We show that the best solution is achieved when each transmitter is assigned to the closest source, each receiver is assigned to the closest destination, and coherent beamforming is performed towards the best selection of receiver.
\end{itemize}
Both analytical and simulation results show significant power gains of a coherent system than a non-coherent system and demonstrate the feasibility of the proposed multi-layer network optimization solution using single-antenna communications.

The rest of the paper is organized as follows.
Section~\ref{sec:related} discusses the related work.
Section~\ref{sec:model} describes the system architecture.
Section~\ref{sec:require} analyzes the requirements for coherent communications.
Section~\ref{sec:beam} designs the protocol for coherent beamforming from a set of transmitters to a set of receivers.
Section~\ref{sec:cluster} designs the network formation protocol that determines a set of transmitters and a set of receivers for each coherent communication group and the beamforming protocol for each group.
Section~\ref{sec:conclusion} concludes the paper.
Appendix provides proofs of Theorems~\ref{theorem:requirement}--\ref{theorem:sirgain} and Corollary~\ref{coro:upperbound}.

\section{Related Work}
\label{sec:related}

Coherent communications were studied mainly for a \emph{single data stream} ($K=1$) with \emph{multiple transmitters} ($N>1$) and a \emph{single receiver} ($M=1$).
Most of these works considered applications in sensor networks \cite{Jayaprakasam17:sensor}.
In \cite{Marques08:MISO}, a coherent beamforming scheme was designed for multiple-input-single-output (MISO) communications  with finite-rate feedback to minimize the transmit power.
In \cite{Nanzer17:MISO}, an open-loop coherent beamforming scheme was designed for MISO communications.
The impact of phase errors on distributed coherent beamforming was analyzed in \cite{Song09:MISO}. For wireless sensor networks, a distributed beamforming scheme  was designed in \cite{Zarifi10:MISO} to improve network connectivity and energy efficiency.
In \cite{Scherber13:MISO}, coherent beamforming was applied for MISO to minimize transmit power, thereby showing the benefits of coherent communications.
Distributed transmit beamforming was designed and implemented with software-defined radio (USRP) \cite{Quitin13:MISO,Hu16:MISO}.
In \cite{Bidigare15},  a distributed transmit beamforming scheme was developed for wideband communications using channel reciprocity and relative calibration. A distributed beamforming scheme was designed in \cite{Gencel1511:MISO} for noise-resilient wideband communications, and the aggregate feedback was considered in \cite{Gencel1506:MISO}.
Coherent beamforming was also studied for radar systems \cite{Deng16:radar}.
We consider a MIMO scheme in this paper.
It can be regarded as a summation of multiple MISO schemes if we do not consider the time to schedule multiple MISO transmissions.
MIMO scheme can transmit data to $M$ receivers in one time slot while the corresponding MISO scheme needs $M$ time slots to transmit data to all $M$ receivers.
As a consequence, the scheme of multiple MISO transmissions would significantly reduce the throughput that can be achieved by MIMO transmissions.

There are many studies on self-organizing networks and they are mostly focused on clustering in wireless networks, e.g., \cite{Shokouhifar17:clustering, Abuarqoub17:clustering, Cenedese17:clustering, Rao17:clustering}.
An adaptive fuzzy clustering protocol was designed in \cite{Shokouhifar17:clustering} to prolong the sensor network lifetime.
In \cite{Abuarqoub17:clustering}, a self-organizing and adaptive clustering solution was developed to improve the sensor network performance such a delivery ratio, delay, and lifetime.
Both a centralized and a distributed clustering algorithm were designed in \cite{Cenedese17:clustering} for sensor networks. Unequal clustering (considering smaller size clusters near to the sink and larger size clusters relatively far away from the sink) was designed in \cite{Rao17:clustering} along with a routing algorithm. This unequal clustering achieves better traffic load balance and thus improves performance such as lifetime, delivery ratio, and convergence rate.
Some related works considered noncoherent beamforming and clustering, e.g., \cite{Papadogiannis08:clustering, Dai14:clustering, Ramamonjison14:clustering}.
\cite{Papadogiannis08:clustering} and \cite{Dai14:clustering} aimed to maximize the total data rate via beamforming and clustering for download traffic.
In \cite{Ramamonjison14:clustering}, an energy efficient solution was designed to meet Quality of Service (QoS)  requirements of users.
These clustering solutions are not designed to optimize the performance of coherent communications.

There are only a few works on joint coherent beamforming and network formation.
\cite{Zheng11:MISO-Clustering} aimed to minimize the transmission power of nodes for sensor networks under the  finite-rate feedback by considering only the MISO case, i.e., a single receiver ($M=1$).
In \cite{Shi18:coherent}, we considered the preliminary study of coherent beamforming and network formation for the case of multiple transmitters and multiple receivers, which includes MISO as a special case.
In this paper, we  consider additional settings, including the impact of Doppler spread, additional channel model (the two-ray model), new beamforming protocols, and extension of joint beamforming and network formation protocols to better characterize the gains of distributed coherent group communications.

\section{System Model}
\label{sec:model}

\begin{figure}
	\centering
	\includegraphics[width=0.8\columnwidth]{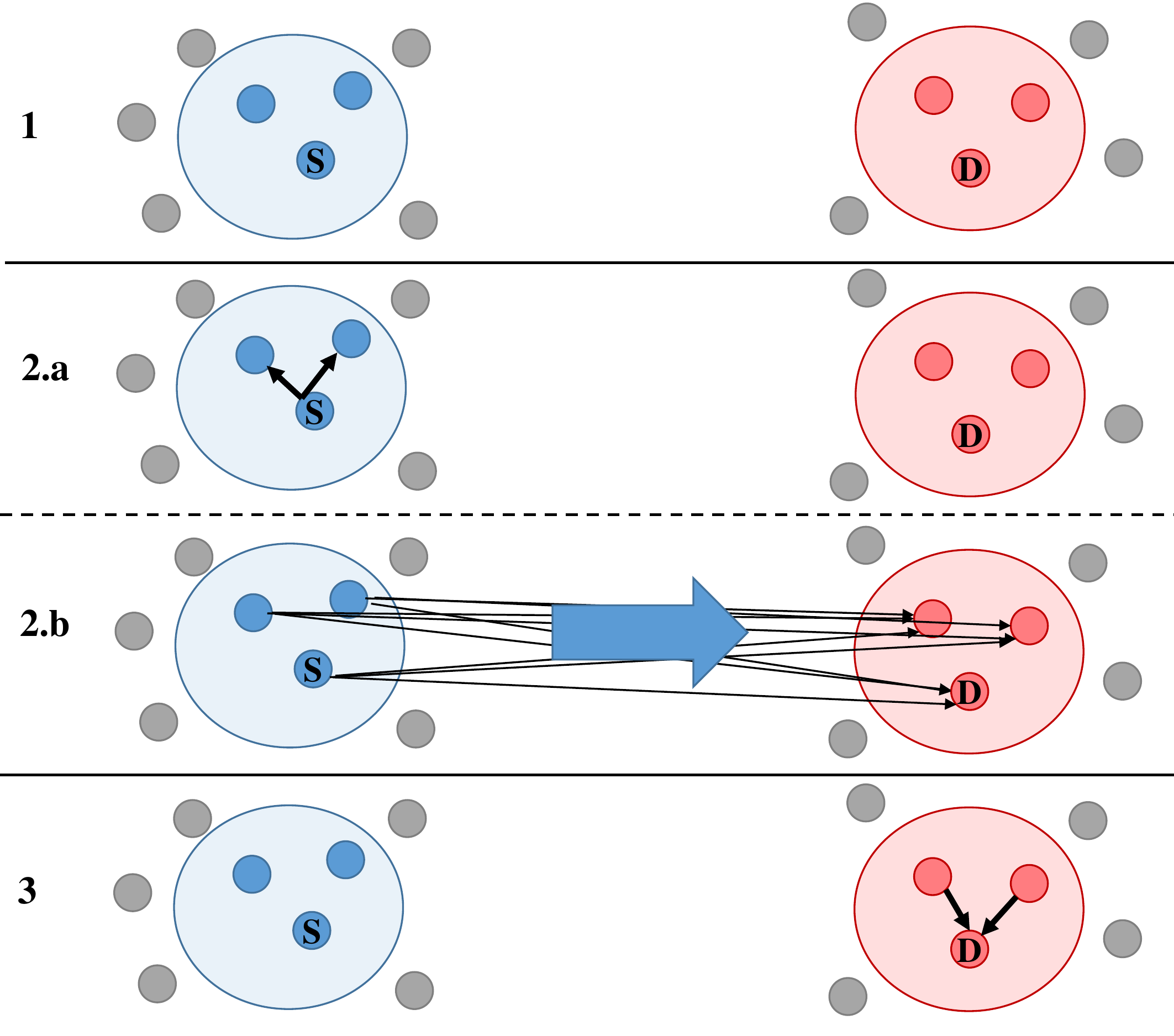}
	\caption{Steps of distributed coherent group communications.}
	\label{fig:phases}
\end{figure}

We consider a coherent communications scenario, where $N$ transmitters transmit data coherently to $M$ receivers.
The distance between transmitter $i$ and receiver $j$ is $d_{ij}$. In Section~\ref{sec:beam}, we consider a single data steam ($K=1$). In Section~\ref{sec:cluster}, we extend the setting to simultaneous operation of $K>1$ data streams such that $N$ transmitters transmit data coherently to $M$ receivers for all data streams. Among $M$ transmitters, there is one source node that has data to transmit to one destination node, which is one of $M$ receivers.
Nodes perform distributed protocols in three main steps as shown in Fig.~\ref{fig:phases}.

\begin{enumerate}
\item \emph{Clustering}: Nodes cluster each other into (disjoint) transmitter and receiver groups.
\item \emph{Transmit protocols}:
\begin{enumerate}
\item \emph{Data exchange in the transmitter group}: The source node broadcasts data
in the transmitter group.

\item \emph{Inter-group communications}: The transmitter group sends data to the receiver group in three steps:
\begin{enumerate}
	\item \emph{channel estimation}: Nodes in the transmitter group send pilot signals and all channels
	are estimated at the receivers.
	\item \emph{feedback}: The channel state information is sent back to nodes in the transmitter group.
	\item \emph{data transmission/beamforming}: All nodes in the transmitter group
	transmit with appropriate amplitudes and phases to the receiver group.
\end{enumerate}
\end{enumerate}
\item \emph{Data exchange in the receiver group}: The destination node collects the received samples from other nodes in the receiver group.
\end{enumerate}

We require the inter-group channels (namely, channels among nodes in the same group) to remain static during the coherent transmission duration. The nodes in the transmitter group exchange the data prior to the coherent transmission duration. The nodes in the receiver group exchange their signals for coherent combining after the coherent transmission from the transmitter group.
Therefore, the intra-group channels do not need to be static during the intra-group data exchange period. To perform coherent communications, it is necessary to collect channel information, e.g., channel gain and phase shift \cite{Pandula07,Pandula11,Alkhateeb14}, between each pair of transmitter and receiver in channel sensing. A simple channel sensing scheme is time multiplexed training, where transmitters transmit the training signal with length $T_t$ sequentially.
Since each transmitter-receiver pair has a different propagation delay $t_{ij}$, the previous transmission may interfere with the current one. Therefore, guard period $T_g$ between transmissions  is used to avoid this type of interference.
The total training time $N(T_t+T_g)$ increases linearly with the number of transmitters, $N$.
This overhead reduces the average data rate
achieved by coherent beamforming over all three steps.

\section{Coherent Communications Requirement}
\label{sec:require}

\subsection{Synchronization Requirement}

For coherent communications, it would be the best if we can synchronize all transmitters and receivers. However, this may be a challenging task, especially when transmitters and receivers are far away from each other. It turns out that such a strict requirement is not needed. Instead, the only requirement to support coherent communications is to synchronize transmitters, as stated in Theorem~\ref{theorem:requirement}.
The proof of Theorem~\ref{theorem:requirement} is given in Appendix~A.

\begin{theorem}
\label{theorem:requirement}
To achieve coherent beamforming at a receiver, the only requirement for synchronization is that transmitters are synchronized.
\end{theorem}

Theorem~\ref{theorem:requirement} shows that there is no need to synchronize transmitters and receivers together for coherent communications. Further, it is easy to see that phase coherence at a receiver can be achieved without the consideration of other receivers. Thus, synchronization among receivers is also not needed.
The problem of synchronization between transmitters has been extensively studied, see e.g., \cite{Peiffer16:syn,Quitin12:syn}, and thus in this paper we assume that such synchronization is available.

\begin{figure}
  \centering
  \includegraphics[width=1.0\columnwidth]{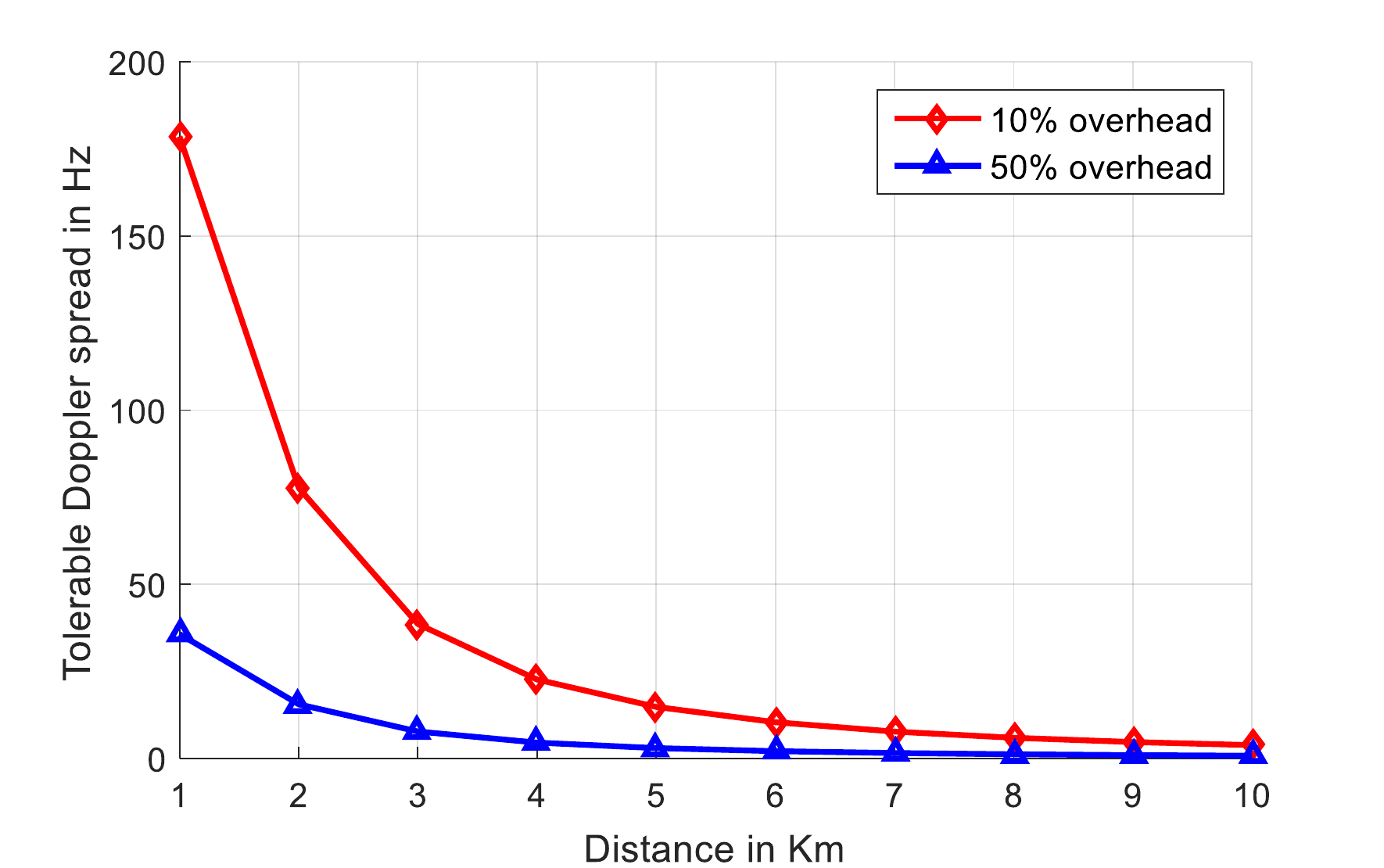}
  \caption{Doppler spread vs. range.}
  \label{fig:doppler}
\end{figure}

\subsection{Doppler Spread Tolerance}

We characterize the \emph{Doppler spread} that can be tolerated to transmit coherently, accounting for channel estimation, quantization, and feedback. Doppler spread and coherence time are inversely proportional. Doppler spread measures the spectral broadening due to the rate of channel changes.

We use the following parameters to assess the tolerable Doppler spread.
Transmitter power is $P_t = 10$ dBm, receiver noise figure is $N_r = 3$ dBm, noise floor is $N_0 = -174$ dBm, signaling bandwidth is $W=1$ MHz, distance $D$ between groups varies from $1$km to $10$km, number of nodes within a group is $N=10$, center frequency is $f_c=2.4$GHz, and speed of light is $c=3 \cdot 10^8$m/s.
Free-space propagation is used to compute the path loss.
The SNR required for accurate channel estimation is $\gamma_t=20$ dBm and the SNR required for channel state feedback to be received successfully at the transmitter group is $\gamma_f=10$ dBm.
Number of bits for each channel estimate is $B_c=16$, number of bits for general header is $B_h=10$, and overhead percentage is selected as either $F_o = 10\%$ or $50\%$ (remaining time is dedicated to data transmission using beamforming).

The calculation of tolerable Doppler spread is as follows. For wavelength $\lambda = \frac{c}{f_c}$, received power in dBm is $P_r = P_t + 20 \log \frac{\lambda}{4 \pi D}$ and received SNR in dBm is $\gamma = P_r - 10 \log(W) - N_r - N_0$.
Then the processing gain needed for training in dBm is $g_t = \gamma_t - \gamma$ and the processing gain needed for feedback in dBm is $g_f = \gamma_f - \gamma$.
Number of training bits is $B_t = \max\{ 10^{g_t/10},N \}$ and the
number of total feedback bits is $B_f = \max\{ 10^{g_f/10},1 \} N B_c + B_h N$. Then total overhead (in bits) is given by $B_o = B_t + B_f$. This overhead is translated to time duration $T_o = \frac{B_o}{W} + \frac{3D}{c}$ (by accounting for the three-step transmission of pilot signal, feedback, and data). Then the minimum coherence time is given by $T_c = \frac{T_o}{F_o}$ when overhead percentage is $F_o$. From $T_c $, tolerable Doppler spread is computed as $S=\frac{1}{T_c}$.
Figure~\ref{fig:doppler} shows that the tolerable Doppler spread decreases as the distance between transmitter and receiver groups increases.

\section{Coherent Beamforming}
\label{sec:beam}

After channel gain and phase shift are obtained between any transmitter and receiver, and synchronization is achieved among transmitters, we now present the coherent beamforming protocol to maximize the power gain at receivers.

\subsection{Coherent Communication and Benchmark Scenarios}

We consider a coherent communication scenario for one data stream from transmitter $1$ to receiver $1$.
There are $N-1$ additional transmitters and $M-1$ additional receivers that also participate in coherent communications.
Suppose transmitters use a sine wave for the signal with the same signal amplitude $A$ and signal period $T$, i.e., signal from transmitter $n$ is $A \sin⁡(\frac{2 \pi}{T} t +\theta_n )$, where $\theta_n$ is the initial phase value that will be determined by a coherent beamforming protocol.
Denote channel gain and phase shift from transmitter $n$ to receiver $m$ as $h_{nm}^2$ and $\theta_{nm}$, respectively, which are obtained in channel estimation.
Thus, receiver $m$ receives the signal $A h_{nm} \sin⁡(\frac{2 \pi}{T} t+ \theta_n+ \theta_{nm})$ from transmitter $n$.
The total received energy during a period at all receivers is
\begin{eqnarray*}
E_{NM} = \sum_{m=1}^M \int_{t=0}^T \left( \sum_{n=1}^N A h_{nm} \sin \left( \frac{2 \pi}{T} t+ \theta_n+ \theta_{nm} \right) \right)^2 dt .
\end{eqnarray*}

We define the benchmark scenario as the case of $N=1$ and $M=1$ (i.e., point-to-point), where coherent communication is not applied. The received energy during a period is
\begin{eqnarray*}
E_{11} = \int_{t=0}^T \left( Ah_{11} \sin \left( \frac{2 \pi}{T} t+ \theta_1 \right) \right)^2 dt = \frac{A^2 h_{11}^2 T}{2} \; .
\end{eqnarray*}

\subsection{Coherent Communication Gain}

We define the ratio of the total received power under the coherent communication scenario and the received power $\frac{A^2 h_{11}^2 T}{2}$ under the benchmark scenario as the coherent communication gain $G(\bm{\theta})$ that is a function of beamforming parameters $\bm{\theta}$ (where the $i$th entry of vector $\bm{\theta}$ is the phase angle $\theta_i$ of node $i$). Theorem~\ref{theorem:gain} presents the coherent communication gain.
The proof of Theorem~\ref{theorem:gain} is given in Appendix~B.

\begin{theorem}
	\label{theorem:gain}
The power gain of coherent communications for one data stream with $N$ transmitters and $M$ receivers is
\begin{eqnarray}
\label{eq:gainnm}
G(\bm{\theta}) = \frac{1}{h_{11}^2} \sum_{m=1}^M \beta_m(\bm{\theta}) \; ,
\end{eqnarray}
where $\beta_m(\bm{\theta})$ is given by
\begin{eqnarray*}
\beta_m(\bm{\theta}) & = &  \left( \sum_{n=1}^N h_{nm} \cos(\theta_n + \theta_{nm}) \right)^2 + \\ &&
\left( \sum_{n=1}^N h_{nm} \sin(\theta_n + \theta_{nm}) \right)^2 \; .
\end{eqnarray*}
\end{theorem}

The upper bound on the coherent communication gain is stated in Corollary~\ref{coro:upperbound}.
Its proof is given in Appendix~C.

\begin{corollary}
	\label{coro:upperbound}
$G(\bm{\theta})$ has an upper bound $G_{\text{UB}} = N^2 M$ if $h_{nm}^2 \approx h_{11}^2$ for $m=1, \cdots, M$ and $n=1, \cdots, N$.
\end{corollary}

Note that we assume the same transmit power at each transmitter and thus we define $A$ as the signal amplitude of a transmitter.
For the general case of different transmit powers, we can define $A_n$ as the signal amplitude of transmitter $n$ and replace each $h_{nm}$ by $A_n h_{nm}$.
Then the result in Theorem~\ref{theorem:gain} is changed to
\begin{eqnarray*}
G(\bm{\theta}) &=& \frac{1}{A_1^2 h_{11}^2} \sum_{m=1}^M \beta_m(\bm{\theta}) \; , \\
\beta_m(\bm{\theta}) &=& \left( \sum_{n=1}^N A_n h_{nm} \cos(\theta_n + \theta_{nm}) \right)^2 + \\
&& \left( \sum_{n=1}^N A_n h_{nm} \sin(\theta_n + \theta_{nm}) \right)^2 \; .
\end{eqnarray*}
The condition in Corollary~\ref{coro:upperbound} is changed to $A_n^2 h_{nm}^2 \approx A_1^2 h_{11}^2$ for $m=1, \cdots, M$ and $n=1, \cdots, N$.

\begin{figure}
  \centering
  \includegraphics[width=1.0\columnwidth]{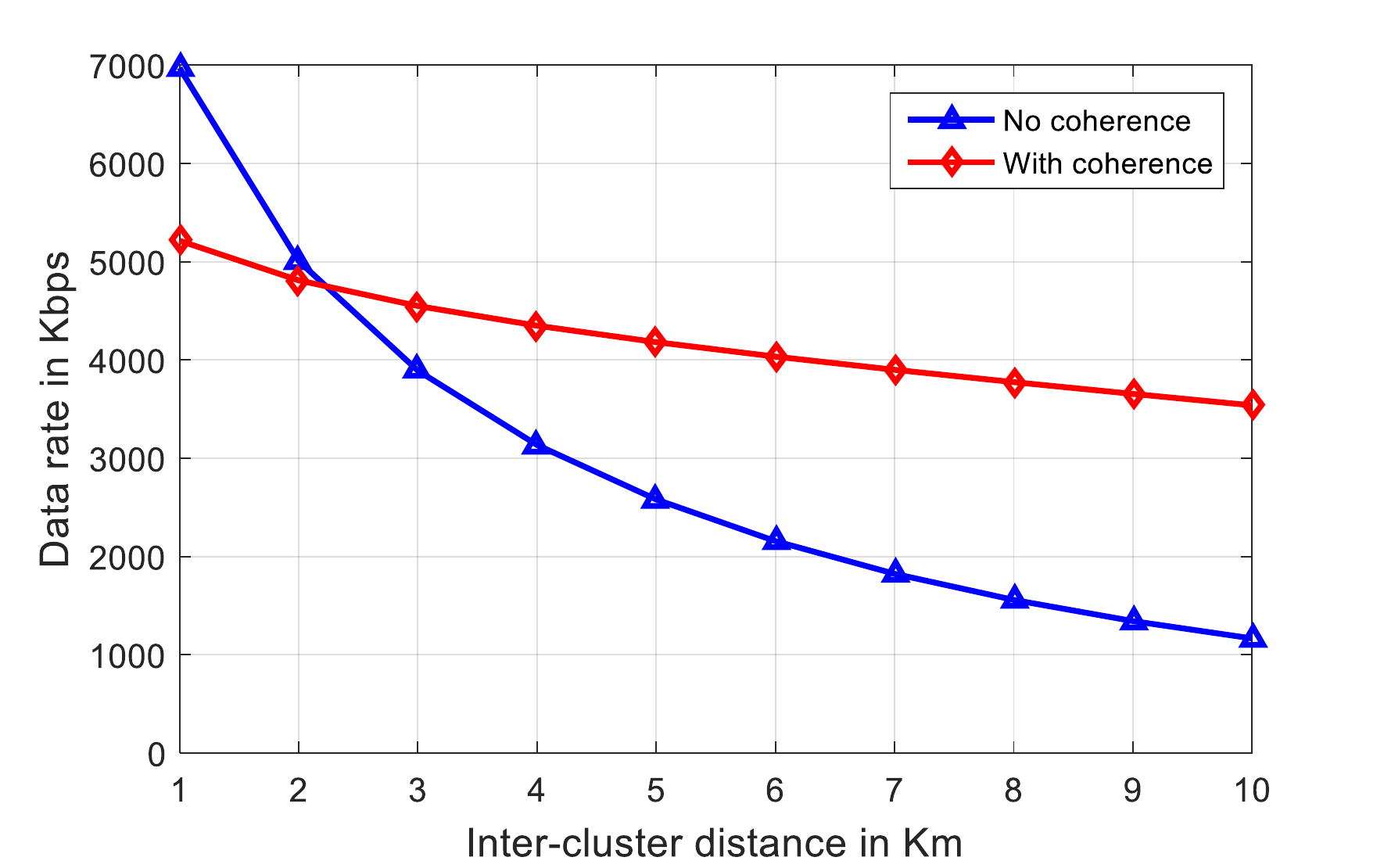}
  \caption{Coherent communications performance.}
  \label{fig:coherent}
\end{figure}

Note that this formulation can be extended to include interference effects and provide the capability of distributed interference cancellation (such as studied in \cite{Ponnaluri2017}).
We compare the improvement in data rate relative to a long-haul point-to-point link without coherent communications (such as studied in \cite{Ponnaluri15,Erpek18:MIMO, Erpek19}). We set the coherence time to be $100$ milliseconds.
We assume the optimum beamforming gain of $N^3$ with $N$ nodes in transmitter and receiver groups ($M=N$).
Figure~\ref{fig:coherent} shows the results. Note that there is a threshold on the inter-group distance such that if the inter-group distance is larger than this threshold, coherent communications are better than point-to-point transmissions.
Point-to-point transmissions can have larger data rate than coherent communications if the inter-group distance is less than this threshold, since the coherent communication gain is not large enough to compensate the coherent communication overhead, namely the time to exchange data in the transmitter/receiver group plus the time to collect more channel information.

\subsection{Coherent Beamforming Protocols}

The optimization problem for coherent beamforming is \begin{eqnarray}
\max & G(\bm{\theta}) \label{eq:opt1} \\
\mbox{subject to} & (\ref{eq:gainnm}) \nonumber \\
\mbox{variables:} & \bm{\theta} = \{\theta_n, 0 \le \theta_n \le 2\pi, 1 \le n \le N \} \; . \nonumber
\end{eqnarray}
This optimization problem can be easily solved if $M=1$.
The optimal solution is not unique and one such solution is $\theta_1 = 0$ and $\theta_n = \theta_1 + \theta_{11} - \theta_{n1}$ for any $n \ge 2$, which achieves
phase coherence at receiver $1$.
This case is called as multiple-input-single-output (MISO) and was widely studied (see e.g., \cite{Marques08:MISO,Song09:MISO,Zarifi10:MISO,Scherber13:MISO,Quitin13:MISO,Bidigare15,Gencel1511:MISO,Gencel1506:MISO,Zheng11:MISO-Clustering}).
However, if $M > 1$, there is no solution to achieve phase coherence at all receivers in general.
The optimization problem is not convex and thus an optimal solution may not be found in polynomial time.

We design two coherent beamforming protocols for this optimization problem to maximize the coherent communications gain $G(\bm{\theta})$.
A \emph{sequential fixing} (SF) protocol to adjust the phase angle $\theta_n$ values is given in Algorithm~\ref{alg:sf}.
Note that in Step 2, we can set $\theta_1=0$ because if there is an optimal solution with $\theta_1 \ne 0$, we can always construct another solution by letting $\hat \theta_n=\theta_n- \theta_1$ for $n=1, \cdots,N$.
It is easy to verify that the constructed solution has the same coherent communication gain and $\hat \theta_1=0$.
Thus, setting $\theta_1=0$ can reduce the algorithm complexity without losing the optimality.

\begin{algorithm}[t]
    \caption{The SF protocol.}
    \label{alg:sf}
    \begin{algorithmic}[1]
        \STATE Sort transmitters based on $\sum_{m=1}^M h_{nm}^2$, i.e., the expected value for received power from a transmitter to all receivers. To simplify discussion, we assume
    $$\sum_{m=1}^M h_{1m}^2 \ge \cdots \ge \sum_{m=1}^M h_{Nm}^2 \; .$$

        \STATE Set $\theta_1=0$.

        \FOR{$n=2, \cdots,N$}
            \STATE Choose $\theta_n$ to maximize the coherent communication gain $G(\bm{\theta})$ by $n$ transmitters
$\sum_{m=1}^M ((\sum_{i=1}^n h_{im} \cos⁡(\theta_i+ \theta_{im}))^2+(\sum_{i=1}^n h_{im} \sin⁡(\theta_i+ \theta_{im}))^2 ) /h_{sd}^2$,
where $\theta_1, \cdots, \theta_{n-1}$ are already determined and $h_{sd}^2$ is the channel gain between the source and the destination.
        \ENDFOR
    \end{algorithmic}
\end{algorithm}

\begin{algorithm}[t]
    \caption{The IO protocol.}
    \label{alg:io}
    \begin{algorithmic}[1]
        \STATE Sort transmitters based on $\sum_{m=1}^M h_{nm}^2$. To simplify discussion, we assume
$$\sum_{m=1}^M h_{1m}^2 \ge \cdots \ge \sum_{m=1}^M h_{Nm}^2 \; .$$

        \STATE Set $\theta_1=0$.

        \FOR{$n=2, \cdots,N$}
            \STATE Choose $\theta_n$ to maximize coherent communication gain $G(\bm{\theta})$ by $n$ transmitters
$\sum_{m=1}^M ((\sum_{i=1}^n h_{im} \cos⁡(\theta_i+ \theta_{im}))^2+(\sum_{i=1}^n h_{im} \sin⁡(\theta_i+ \theta_{im}))^2 ) /h_{sd}^2$.
        \ENDFOR

        \STATE Set Improved = 1.

        \WHILE{Improved == 1}
            \STATE Set Improved = 0.

            \FOR{$n=1, \cdots,N$}
                \STATE Choose $\theta_n$ to maximize the coherent communication gain $G(\bm{\theta}) =
        \sum_{m=1}^M ((\sum_{i=1}^N h_{im} \cos⁡(\theta_i+ \theta_{im}))^2+(\sum_{i=1}^N h_{im} \sin⁡(\theta_i+ \theta_{im}))^2 ) /h_{sd}^2$,
        where $\theta_i$ values, $i \ne n$, are unchanged.

                \STATE If $\theta_n$ is updated, set Improved = 1.
            \ENDFOR
        \ENDWHILE
    \end{algorithmic}
\end{algorithm}

Next, we introduce the \emph{iterative optimization} (IO) protocol to iteratively optimize the phase angle $\theta_n$ values. The algorithmic steps of the IO protocol are given in Algorithm~\ref{alg:io}.
The first three steps in the IO protocol are the same as in the SF protocol and the next two steps try to improve the coherent communication gain $G(\bm{\theta})$. Thus, the IO protocol always achieves better performance than the SF protocol but it has higher complexity.

The designed distributed protocols can also be implemented as centralized algorithms (by assuming all information available at one node).  The performance and computational complexity are the same for the distributed protocol and its centralized version. However, the distributed protocol incurs additional communication overhead (exchanging intermediate results) that is not present in its centralized version.

For comparison purposes, we also design other protocols.
\begin{enumerate}
\item \emph{Random beamforming} (RB) protocol. Randomly choose $\theta_n$ values corresponding to no coherent communication.

\item \emph{Random target} (RT) protocol. Choose $\theta_n$ values to achieve phase coherence at a random receiver. If the target receiver is $m$, this can be done by setting $\theta_1=0, \theta_n=\theta_{1m}- \theta_{nm}$ for $n=2, \cdots,N$.

\item \emph{Best target} (BT) protocol. Instead of considering $M$ receivers jointly, we study the coherent communication gain achieved by $N$ transmitters and each receiver, and identify the the best receiver with the largest coherent communication gain. Then we choose $\theta_n$ values to achieve this gain and let all receivers participate in coherent communications.

\item \emph{Exhaustive search} (ES) protocol. Exhaustively search for the best $\theta_n$ values to maximize the coherent communication gain $G(\bm{\theta})$. Given the high complexity of the ES protocol, we implement this protocol for $N \le 3$.
\end{enumerate}
The performance achieved by ES protocol, as well as the upper bound $G_{\text{UB}}$ in Corollary~\ref{coro:upperbound}, are used as benchmarks.

\begin{table}
\caption{Coherent communication gain under Channel Model 1 with $D=1000$ and $r=10$.}
\scriptsize
\centering
\begin{tabular}{c|r|r|r|r|r|r|r}
\multicolumn{1}{c|}{$(N,M)$} & \multicolumn{1}{c|}{RB} & \multicolumn{1}{c|}{RT} & \multicolumn{1}{c|}{BT} & \multicolumn{1}{c|}{SF} & \multicolumn{1}{c|}{IO} & \multicolumn{1}{c|}{ES} & \multicolumn{1}{c}{$G_{\text{UB}}$} \\ \hline \hline
$(1,1)$   & \bf{1}     & \bf{1}      & \bf{1}      & \bf{1}      & \bf{1}      & 1     & 1    \\ \hline
$(1,10)$  & \bf{9.97}  & \bf{9.97}   & \bf{9.97}   & \bf{9.97}   & \bf{9.97}   & 9.97  & 10    \\ \hline
$(3,1)$   & 3.04  & \bf{8.98}   & \bf{8.98}   & \bf{8.98}   & \bf{8.98}   & 8.98  & 9    \\ \hline
$(3,10)$  & 26.82 & 87.27  & 88.71  & \bf{88.73}  & \bf{88.73}  & 88.73 & 90    \\ \hline
$(10,10)$ & 99.82 & 840.17 & 909.13 & 909.71 & \bf{910.47} & -     & 1000
\end{tabular}
\label{table:coherent}
\end{table}

\begin{table}
\caption{Coherent beamforming protocols.}
\scriptsize
\centering
\begin{tabular}{c|c}
RB   & Random beamforming protocol    \\ \hline
RT  & Random target protocol    \\ \hline
BT   & Best target protocol    \\ \hline
SF  & Sequential fixing protocol    \\ \hline
IO & Iterative optimization protocol    \\ \hline
ES & Exhaustive search protocol
\end{tabular}
\label{table:protocol1}
\end{table}

To evaluate the proposed protocols, we consider the following setting. There is one data stream ($K=1$) from transmitter $1$ to receiver $1$ and the distance between them is $D$.
There are $N-1$ additional transmitters and $M-1$ additional receivers that can participate in coherent communications.
The distance between transmitters $i$ and $1$ is no more than $r$ and the distance between receivers $i$ and $1$ is also no more than $r$.
We set $D \gg r$ such that transmitters (and receivers) can collaborate to transmit over a long distance.
Channel gain is modeled by $h^2=d^{-2}$ (Channel Model 1) and phase shift is modeled by $\theta=2 \pi(d/\lambda- \lfloor d/\lambda \rfloor)$, where $d$ is the distance between a transmitter and a receiver, and $\lambda$ is the wavelength.

Table~\ref{table:coherent} shows the coherent communication gain results obtained for $D=1000$m, $r=10$m, and $\lambda=0.125$m (corresponding to $2.4$GHz),
where full names of protocols are listed in Table~\ref{table:protocol1}.
The RB protocol uses a random algorithm and thus we show the average performance in $100$ runs.

Two scenarios with $N=1$ are used to verify the correctness of protocol implementation.
For $N=1$ and $M=10$, we have $G_{\text{UB}} = 10$ under the assumption that all channel gains are the same.
Since this assumption does not hold in simulation, the ES protocol finds that the maximum coherent communication gain $G(\bm{\theta})$ is $9.97$, which is slightly less than $G_{\text{UB}} = 10$.
Due to the same reason, the maximum gain $G(\bm{\theta})$ obtained by the ES protocol for scenario $(N,M)=(3,1)$ or $(3,10)$ is slightly less than $G_{\text{UB}}= 9$ or $90$.

For scenario $(N,M)=(3,1)$, most protocols (except the RB protocol) find the optimal solution with the maximum gain $G(\bm{\theta}) = 8.98$ while the RB protocol cannot.
In fact, the RB protocol is the case of non-coherent communication and thus has the expected gain $G(\bm{\theta}) = NM$.
For scenario $(N,M)=(3,10)$, the SF and IO protocols can find the optimal solution with the maximum gain $G(\bm{\theta}) = 88.73$.
The RT and BT protocols can find near-optimal solutions, while the RB protocol finds a solution with gain $G(\bm{\theta}) = 26.82$, which is far from the optimum.

For scenario $(N,M)=(10,10)$, the ES protocol cannot find a solution due to high complexity and thus the maximum gain is unknown.
Since the IO protocol finds the best solution (with gain $G(\bm{\theta}) = 910.47$) among all protocols and $G_{\text{UB}}= 1000$,
the maximum gain is within $[910.47, 1000]$.
The BT and SF protocols can find solutions with a gain close to the best known gain $G(\bm{\theta}) = 910.47$.
The RB protocol again cannot find a good solution.

The state-of-the-art methods focus on the MISO communications, which corresponds to the case of $M=1$ in Table~\ref{table:coherent}.
For the MISO case, the optimal solution can be achieved.
Most schemes developed in this paper, such as schemes RT to IO in Table~\ref{table:coherent}, can achieve the optimal gain.
For the case of $M>1$, there is no previous method and thus we develop the ES scheme for small $N$ and establish an upper bound $G_{UB}$ for large $N$ as the performance benchmarks.

In summary, the IO protocol always has the best performance among all protocols and achieves the maximum gain $G(\bm{\theta})$ when the ES protocol is applicable.
The BT and SF protocols are also viable in terms of providing near-optimal solutions, while the RB protocol should not be used due to its poor performance.

We obtain more results under different parameters and channel models, when the radius $r$ of transmitter/receiver group is $100$m and the distance $D$ between
these two groups is $1$km or $10$km.
We use realistic channel models:
\begin{enumerate}
\item \emph{Free space model} (Channel Model 2): $h^2= \frac{G_t G_r \lambda^2}{(4 \pi d)^2}$, where $G_t$ and $G_r$ are the transmitter and receiver antenna gains, respectively.

\item \emph{Two-ray model} (Channel Model 3): If $d$ is less than crossover distance $d_c$, $h^2$ is calculated by the free space model, else $h^2= \frac{G_t G_r (h_t h_r )^2}{d^4}$, where $d_c= \frac{4 \pi h_t h_r}{\lambda}$, $h_t$ and $h_r$ are the transmitter and receiver antenna heights, respectively.
\end{enumerate}
We set $G_t=G_r=1, \lambda=0.125$ and $h_t=h_r=0.5$ in simulation and obtain results in Table~\ref{table:free1}--Table~\ref{table:tworay2}
for $r=100$m and $D =1$km or $10$km.
The column of $G_{\text{UB}}$ shows the upper bound $N^2 M$.
The ES protocol can only find the optimal solution for small $N$, i.e., it cannot find a solution for $N=10$.
Due to larger $r$ ($r=100$m), the gap from the condition $h_{nm}^2 \approx h_{11}^2$ for $m=1,\cdots,M$ and $n=1,\cdots,N$ in Corollary~\ref{coro:upperbound}  is larger than the case when $r=10$m.
As a result, the upper bound $G_{\text{UB}}$ for $r =100$m obtained by Corollary~\ref{coro:upperbound} is not as tight (not as close to the optimal value by the ES protocol) as in the case when $r=10$m.
The IO protocol has the largest power gain $G(\bm{\theta})$ but this gain is not close to $G_{\text{UB}}$.
For $N=3$, the IO protocol finds a solution with similar performance as that achieved by the ES protocol, which indicates that this solution is near optimal.

\begin{table}[t!]
\caption{Coherent communication gain under Channel Model 2 with $D=1000$ and $r=100$.}
\label{table:free1}
\scriptsize
\centering
\begin{tabular}{c|r|r|r|r|r|r|r}
\multicolumn{1}{c|}{$(N,M)$} & \multicolumn{1}{c|}{RB} & \multicolumn{1}{c|}{RT} & \multicolumn{1}{c|}{BT} & \multicolumn{1}{c|}{SF} & \multicolumn{1}{c|}{IO} & \multicolumn{1}{c|}{ES} & \multicolumn{1}{c}{$G_{\text{UB}}$} \\ \hline \hline
$(3,10)$  & 34.91  & 36.84  & 41.56  & 42.87  & \bf{43.59}  & 43.59 & 90 \\ \hline
$(10,10)$ & 106.88 & 177.68 & 195.74 & 223.24 & \bf{257.74} & -     & 1000
\end{tabular}
\end{table}

\begin{table}[t!]
\caption{Coherent communication gain under Channel Model 3 with $D=1000$ and $r=100$.}
\label{table:tworay1}
\scriptsize
\centering
\begin{tabular}{c|r|r|r|r|r|r|r}
\multicolumn{1}{c|}{$(N,M)$} & \multicolumn{1}{c|}{RB} & \multicolumn{1}{c|}{RT} & \multicolumn{1}{c|}{BT} & \multicolumn{1}{c|}{SF} & \multicolumn{1}{c|}{IO} & \multicolumn{1}{c|}{ES} & \multicolumn{1}{c}{$G_{\text{UB}}$} \\ \hline \hline
$(3,10)$  & 42.84  & 42.84  & 48.81  & 49.91  & \bf{50.85}  & 50.85 & 90 \\ \hline
$(10,10)$ & 110.27 & 186.30 & 202.40 & 229.01 & \bf{266.46} & -     & 1000
\end{tabular}
\end{table}

\begin{table} [t!]
\caption{Coherent communication gain under Channel Model 2 with $D=10000$ and $r=100$.}
\label{table:free2}
\scriptsize
\centering
\begin{tabular}{c|r|r|r|r|r|r|r}
\multicolumn{1}{c|}{$(N,M)$} & \multicolumn{1}{c|}{RB} & \multicolumn{1}{c|}{RT} & \multicolumn{1}{c|}{BT} & \multicolumn{1}{c|}{SF} & \multicolumn{1}{c|}{IO} & \multicolumn{1}{c|}{ES} & \multicolumn{1}{c}{$G_{\text{UB}}$} \\ \hline \hline
$(3,10)$  & 29.40  & 38.36  & 49.47  & 50.51  & \bf{50.63}  & 50.63 & 90 \\ \hline
$(10,10)$ & 101.38 & 228.66 & 295.63 & 295.63 & \bf{321.06} & -     & 1000
\end{tabular}
\end{table}

\begin{table}[t!]
\caption{Coherent communication gain under Channel Model 3 with $D=10000$ and $r=100$.}
\label{table:tworay2}
\scriptsize
\centering
\begin{tabular}{c|r|r|r|r|r|r|r}
\multicolumn{1}{c|}{$(N,M)$} & \multicolumn{1}{c|}{RB} & \multicolumn{1}{c|}{RT} & \multicolumn{1}{c|}{BT} & \multicolumn{1}{c|}{SF} & \multicolumn{1}{c|}{IO} & \multicolumn{1}{c|}{ES} & \multicolumn{1}{c}{$G_{\text{UB}}$} \\ \hline \hline
$(3,10)$  & 29.66 & 38.69  & 49.89  & 50.94  & \bf{51.06}  & 51.06 & 90 \\ \hline
$(10,10)$ & 98.62 & 226.54 & 292.12 & 291.76 & \bf{317.08} & -     & 1000
\end{tabular}
\end{table}

\begin{figure}[t]
  \centering
  \includegraphics[width=0.8\columnwidth]{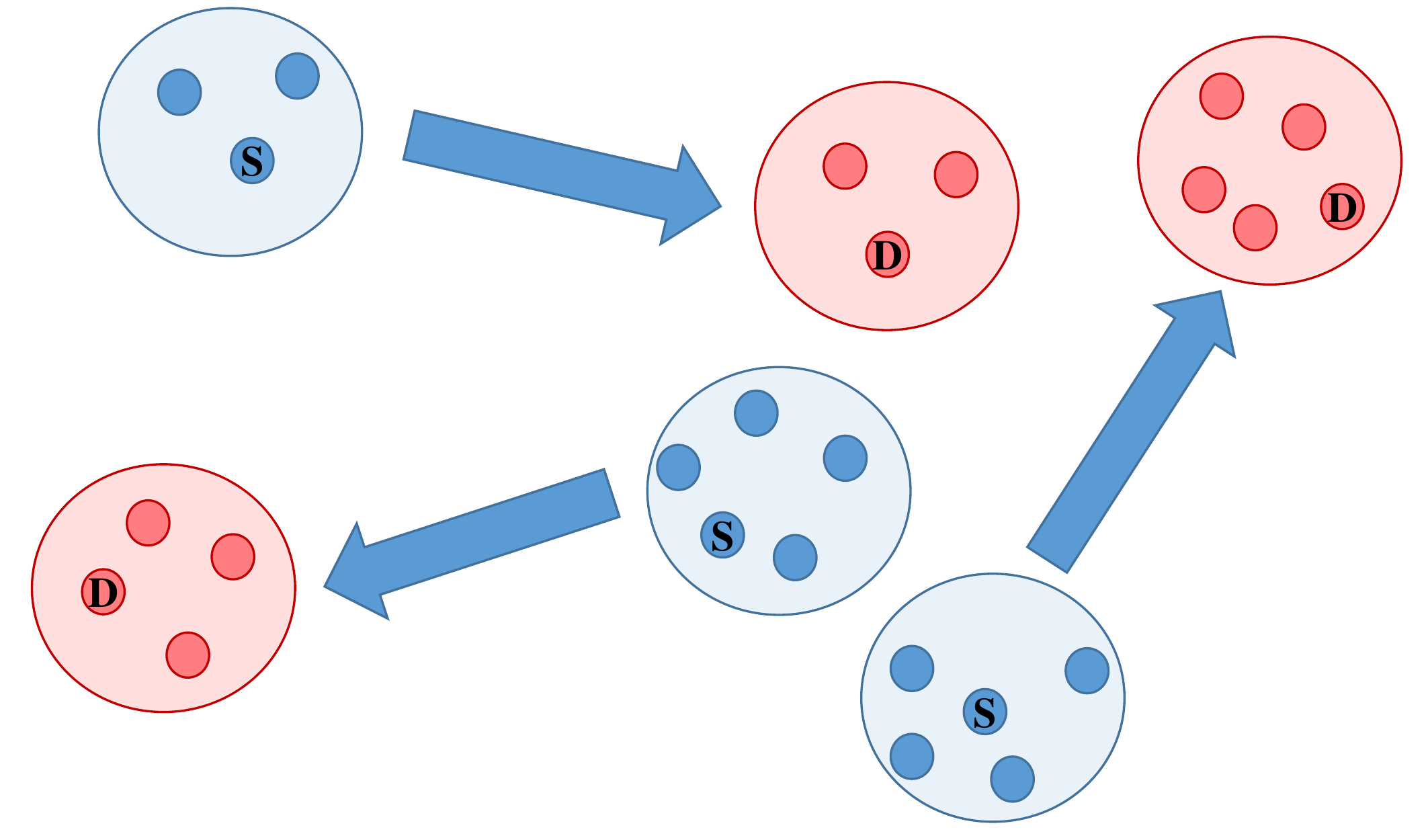}
  \caption{Coherent communications for multiple data streams.}
  \label{fig:clustering}
\end{figure}

\section{Network Formation and Beamforming in Self-organizing Wireless Networks}
\label{sec:cluster}

When there are $K>1$ data streams (see Fig.~\ref{fig:clustering}), each with its own source-destination pair, we need to assign each transmitter and each receiver to a source and a destination, respectively. There are three
\emph{network formation} approaches.
\begin{itemize}
\item Approach 1: Use all nodes in the transmitter group together for beamforming. The problem is that the cost/overhead for synchronization and coordination for beamforming will be very high (increasing with the number of nodes).
	
\item Approach 2: Remove some nodes (potentially distant ones) from the transmitter/receiver group. The problem is that sets of nodes to be removed for different data streams (source-destination pairs) may be different.
	
\item Approach 3: Dynamically group transmitters and receivers to different transmitter and receiver groups, respectively, depending on the data stream requirements.

\end{itemize}

We follow Approach 3 since it selects suitable transmitters and receivers for different data streams and accounts for factors such as traffic backlogs at transmitters and priorities and QoS requirements of data streams.

\subsection{Coherent Communication Gain with Network Formation}

Network formation affects the coherent communication gain $G(\bm{\theta})$. Note that so far we used the power gain as the performance measure for $K=1$.
However, when $K>1$, the power gain cannot capture the effect of interference from other groups. Thus, we use the SIR gain as the measure. There are $K$ data streams, where data stream $k$ is from transmitter $s_k$ to receiver $d_k$.
There are $N$ potential transmitters and $M$ potential receivers for each of these $K$ data streams.
The signal amplitude $A$ and the signal period $T$ are the same for all transmitters.
Denote the channel gain and the phase shift from transmitter $n$ to receiver $m$ as $h_{nm}^2$ and $\theta_{nm}$, respectively.
The signal from transmitter $n$ to receiver $m$ is $Ah_{nm}  \sin{\left(\frac{2 \pi}{T} t+ \theta_n+ \theta_{nm} \right)}$, where transmitter $n$ can tune $\theta_n$.
Denote $\mathcal{T}_k$ and $\mathcal{R}_k$ as the set of transmitters (including $s_k$) and receivers (including $d_k$) assigned to data stream $k$, respectively.
The total received energy for data stream $k$ is $\sum_{m \in \mathcal{R}_k} \int_{t=0}^T \left(\sum_{n \in \mathcal{T}_k} Ah_{nm} \sin{\left(\frac{2 \pi}{T} t+ \theta_n+ \theta_{nm}\right)}\right)^2 dt$.

Theorem~\ref{theorem:sirgain} presents the SIR gain result. The proof of Theorem~\ref{theorem:sirgain} is given in Appendix~D.

\begin{theorem}
	\label{theorem:sirgain}
The SIR gain
of coherent communications for data stream $k$ with transmitter set $\mathcal{T}_k$ and receiver set $\mathcal{R}_k$ is
\begin{eqnarray}
G_k(\bm{\mathcal{T}}, \bm{\mathcal{R}}, \bm{\theta}) &=&
\frac{\sum_{1 \le n \le K}^{n \ne k} h_{s_n d_k}^2 \rho_k(\bm{\theta})}{h_{s_k d_k}^2 \sum_{m \in \mathcal{R}_k} \sum_{n \in \mathcal{T}_l}^{l \ne k} h_{nm}^2} \; , \label{eq:gaink}
\end{eqnarray}
where $\bm{\mathcal{T}}$ and $\bm{\mathcal{R}}$ are sets of all $\mathcal{T}_k$ and $\mathcal{R}_k$, respectively, and
$\rho_k(\bm{\theta})$ is given by
\begin{eqnarray}
\rho_k(\bm{\theta}) &=& \sum_{m \in \mathcal{R}_k} \left( \left( \sum_{n \in \mathcal{T}_k} h_{nm} \cos{⁡(\theta_n+\theta_{nm})} \right)^2
\right . \nonumber \\
&& \left.
+ \left( \sum_{n \in \mathcal{T}_k} h_{nm} \sin{⁡(\theta_n+ \theta_{nm})} \right)^2 \right) \; . \label{eq:rho}
\end{eqnarray}
\end{theorem}

For the case of different transmit powers, we define $A_n$ as the signal amplitude of transmitter $n$ and replace each $A h_{nm}$ by $A_n h_{nm}$.
The result in Theorem~\ref{theorem:sirgain} is changed to
\begin{eqnarray*}
G_k(\bm{\mathcal{T}}, \bm{\mathcal{R}}, \bm{\theta}) &=&
\frac{\sum_{1 \le n \le K}^{n \ne k} A_{s_n}^2 h_{s_n d_k}^2 \rho_k(\bm{\theta})}{A_{s_k}^2 h_{s_k d_k}^2 \sum_{m \in \mathcal{R}_k} \sum_{n \in \mathcal{T}_l}^{l \ne k} A_n^2 h_{nm}^2} \; , \\
\rho_k(\bm{\theta}) &=& \sum_{m \in \mathcal{R}_k} \left( \left( \sum_{n \in \mathcal{T}_k} A_n h_{nm} \cos{⁡(\theta_n+\theta_{nm})} \right)^ 2
\right. \\
&& \left. +\left( \sum_{n \in \mathcal{T}_k} A_n h_{nm} \sin{⁡(\theta_n+ \theta_{nm})} \right)^2 \right) \; .
\end{eqnarray*}

\begin{table}
\caption{Network formation and coherent beamforming protocols.}
\scriptsize
\centering
\begin{tabular}{c|c}
RRB   & Random network formation and random beamforming protocol    \\ \hline
RRT  & Random network formation and random target protocol    \\ \hline
RBT   & Random network formation and best target protocol    \\ \hline
DRB  & Distance-based network formation and random beamforming protocol    \\ \hline
DRT & Distance-based network formation and random target protocol    \\ \hline
DBT & Distance-based network formation and best target protocol
\end{tabular}
\label{table:protocol2}
\end{table}

\begin{table}
\caption{Coherent communication gain under Channel Model 2 with $D=1000$ and $r=100$.}
\label{table:free3}
\small
\centering
\begin{tabular}{c|r|r|r|r|r|r}
$(N,M)$   & RRB  & RRT  & RBT  & DRB  & DRT  & DBT \\ \hline \hline
$(3,10)$  & 2.34 & 2.39 & 2.39 & \bf{2.50} & \bf{2.50} & \bf{2.50} \\ \hline
$(10,10)$ & 2.98 & 5.74 & 6.74 & 1.31 & 7.94 & \bf{8.31}
\end{tabular}
\end{table}

\subsection{Joint Beamforming and Network Formation Protocols}

The joint coherent beamforming and network formation problem can be formulated as
\begin{eqnarray}
\max & f\left( \{G_k( \bm{\mathcal{T}}, \bm{\mathcal{R}}, \bm{\theta}) \}_{k=1,2, \cdots} \right) \label{eq:opt2} \\
\mbox{subject to} & (\ref{eq:gaink}), (\ref{eq:rho}) \nonumber \\
& s_k \in \mathcal{T}_k, d_k \in \mathcal{R}_k & (1 \le k \le K) \nonumber \\
& \mathcal{T}_i \bigcap \mathcal{T}_j = \emptyset, \mathcal{R}_i \bigcap \mathcal{R}_j = \emptyset & (1 \le i \ne j \le K) \nonumber \\
& 0 \le \theta_n \le 2 \pi & (1 \le n \le N) \nonumber \\
\mbox{variables} & \mathcal{T}_k, \mathcal{R}_k, \theta_n \; . \nonumber
\end{eqnarray}
In (\ref{eq:opt2}), $f(\cdot)$ captures the optimization objective and the trade-off among all data streams.
Different functions, e.g., minimum or average, can be used for the function $f(\cdot)$. In simulation results, we use
\begin{equation}
	f\left(\{G_k ( \bm{\mathcal{T}}, \bm{\mathcal{R}}, \bm{\theta}) )\}_{k=1,2,\cdots} \right) = \min_k \{G_k ( \bm{\mathcal{T}}, \bm{\mathcal{R}}, \bm{\theta}) )\} \; . \nonumber
\end{equation}
The problem is to choose the optimal values of $\mathcal{T}_k, \mathcal{R}_k$ and $\theta_n$ according to the optimization problem (\ref{eq:opt2}).
The joint optimization protocol includes network formation policy and beamforming policy.
The network formation policy has a solution with integer variables to describe the assignment a transmitter (receiver) to one of the transmitter (receiver) set.
Such a mixed-integer optimization problem is NP-hard in general.
The complexity to obtain optimal solution is high while a tight upper bound is not obvious.
Thus, to solve this problem, we consider two policies for network formation:
\begin{enumerate}
\item random network formation,
\item distance-based network formation that assigns a transmitter (receiver) to the closest source (destination).
\end{enumerate}
We consider three policies for beamforming within a group:
\begin{enumerate}
\item random beamforming,
\item random target (perform coherent beamforming towards a random receiver),
\item best target (perform coherent beamforming towards the best receiver).
\end{enumerate}
The combination of the above policies results in $2 \times 3=6$ protocols, namely ``Random network formation and Random Beamforming" (RRB), ``Random network formation and Random Target" (RRT),
``Random network formation and Best Target" (RBT), ``Distance-based network formation and Random Beamforming" (DRB),
``Distance-based network formation and Random Target" (DRT),
and ``Distance-based network formation and Best Target" (DBT), respectively.
We list these protocol names in Table~\ref{table:protocol2}.
Simulation results are shown in Table~\ref{table:free3}--Table~\ref{table:tworay4}.
Table~\ref{table:free3} shows results for channel model 2 with distance between transmitter and receiver groups $D=1000$m and group radius $r=100$m.
There are $K=2$ data streams.
If the number of transmitters and receivers $(N,M)$ is $(3,10)$, policies DRB, DRT, and DBT can all achieve the best performance.
But if the number of transmitters $N$ is increased to $10$, only DBT can achieve the best performance among all policies.
We obtain the same conclusion from the other three tables that
DBT achieves the best performance among all policies, highlighting the importance of the distance for network formation and the optimal selection of target receiver for beamforming.

\begin{table}
\caption{Coherent communication gain under Channel Model 3 with $D=1000$ and $r=100$.}
\label{table:tworay3}
\small
\centering
\begin{tabular}{c|r|r|r|r|r|r}
$(N,M)$   & RRB  & RRT  & RBT  & DRB  & DRT  & DBT \\ \hline \hline
$(3,10)$  & 2.26 & 2.32 & 2.32 & \bf{2.52} & \bf{2.52} & \bf{2.52} \\ \hline
$(10,10)$ & 3.21 & 6.13 & 7.21 & 1.23 & \bf{7.67} & \bf{7.67}
\end{tabular}
\end{table}

\begin{table}
\caption{Coherent communication gain under Channel Model 2 with $D=10000$ and $r=100$.}
\label{table:free4}
\small
\centering
\begin{tabular}{c|r|r|r|r|r|r}
$(N,M)$   & RRB  & RRT  & RBT  & DRB  & DRT   & DBT \\ \hline \hline
$(3,10)$  & 2.45 & 2.51 & 2.51 & 2.87 & \bf{4.04}  & \bf{4.04} \\ \hline
$(10,10)$ & 2.57 & 7.52 & 8.69 & 2.62 & 11.19 & \bf{12.35}
\end{tabular}
\end{table}

\begin{table}[t!]
\caption{Coherent communication gain under Channel Model 3 with $D=10000$ and $r=100$.}
\label{table:tworay4}
\small
\centering
\begin{tabular}{c|r|r|r|r|r|r}
$(N,M)$   & RRB  & RRT  & RBT  & DRB  & DRT   & DBT \\ \hline \hline
$(3,10)$  & 2.47 & 2.37 & 2.37 & 2.84 & \bf{4.07}  & \bf{4.07} \\ \hline
$(10,10)$ & 2.60 & 7.24 & 8.31 & 2.89 & 11.20 & \bf{12.38}
\end{tabular}
\end{table}

In the above study, we focus on network formation and coherent beamforming.
We can also consider power control to further improve system performance.
That is, noting that the signal power of one data streams is interference to other data streams, maximum power may not be an optimal solution for optimization problem (\ref{eq:opt2}).
Power control problem has been well studied for SISO links and there are many existing solutions based on nonlinear optimization~\cite{Shi08,Shi09, Lei12}, game theory~\cite{Chen05,Heikkinen06}, metaheuristic~\cite{ElNainay08,Guo11}, etc.
Given the complexity of power control problem, we leave the extension of existing SISO power control algorithms as a future work.

\section{Conclusion}
\label{sec:conclusion}

We presented distributed coherent group communication protocols that consist of coherent beamforming and network formation protocols.
We started with an optimization problem for a single data stream that formulates coherent beamforming for a given pair of transmitter group and receiver group, and obtained a number of policies with near optimal performance in terms of the achieved power gain at the receiver group.
We then considered the joint optimization of coherent beamforming and network formation for multiple data streams to maximize the SIR gain at the receiver group for each coherent communication group.
The network formation protocol forms multiple coherent communication groups for multiple source-destination pairs, where each transmitter group includes a source and each receiver group includes a destination.
Then the beamforming protocol enables transmissions of signals from each transmitter in a group to each receiver in the corresponding receiver group.
The power and SIR gains achieved by our design are higher than those in point-to-point transmissions and bring various advantages such as improvement in communication range, power efficiency, reliability, and throughput.

\section*{Acknowledgement}
We would like to thank Dr. Satya Ponnaluri and Dr. Eric van Doorn for discussions on beamforming and synchronization.

\section*{Appendix A -- Proof of Theorem~\ref{theorem:requirement}}
We first consider a receiver $j$. Suppose that the local time $t$ at transmitter $i$ corresponds to the local time $\hat t$̂ at receiver $j$ and the difference between local times $t_{ij}=t- \hat t$̂ is a fixed number.
If transmitter $i$ sends a signal $A \sin \frac{2 \pi t}{T}$, where $A$ is the amplitude and $T$ is the carrier
period, the received signal at receiver $j$ is $\hat A \sin \left(\frac{2 \pi t}{T} + \frac{2 \pi t_{ij}}{T} +\theta_{ij} \right)$, where $\theta_{ij}$ is the phase shift on link $(i,j)$, and the phase shift is observed as $s_{ij} = \frac{2 \pi t_{ij}}{T} +\theta_{ij}$.
In this case, receiver $j$ can simply ask transmitter $i$ to tune its initial phase value from $0$ to $-s_{ij}$, i.e., transmitter $i$ sends a signal $A \sin⁡ \left(\frac{2 \pi t}{T}-s_{ij} \right)$.
Then the received signal becomes $\hat A \sin⁡ \frac{2 \pi t}{T}$, i.e., with the initial phase value $0$ under the local time at receiver $j$.
If receiver $j$ performs the same operation for each transmitter and each transmitter can tune its signal as requested, phase coherence can be achieved.

\section*{Appendix B -- Proof of Theorem~\ref{theorem:gain}}

We first analyze the received power at a receiver $m$ under the coherent communication scenario. We obtain
\begin{eqnarray*}
&& \hspace{-10mm} E_{NM}(m) = \int_{t=0}^T \left( \sum_{n=1}^N A h_{nm} \sin⁡ \left( \frac{2 \pi}{T} t+ \theta_n+ \theta_{nm} \right) \right)^2 dt \\
&=& A^2 \int_{t=0}^T \left( \sin \frac{2 \pi t}{T} \sum_{n=1}^N h_{nm} \cos⁡(\theta_n+ \theta_{nm})+ \right . \\
&& \left . \cos \frac{2 \pi t}{T} \sum_{n=1}^N h_{nm} \sin⁡(\theta_n+ \theta_{nm}) \right)^2 dt \\
&=& A^2 \left( \sum_{n=1}^N h_{nm} \cos(\theta_n + \theta_{nm}) \right)^2 \int_{t=0}^T \sin^2 \frac{2 \pi t}{T} dt + \\
&& A^2 \left( \sum_{n=1}^N h_{nm} \sin(\theta_n + \theta_{nm}) \right)^2 \int_{t=0}^T \cos⁡^2 \frac{2 \pi t}{T}  dt +
\end{eqnarray*}
\begin{eqnarray*}
&& A^2 \sum_{n=1}^N h_{nm} \cos(\theta_n + \theta_{nm}) \sum_{n=1}^N h_{nm} \sin(\theta_n + \theta_{nm}) \\
&& \cdot \int_{t=0}^T 2 \sin \frac{2 \pi t}{T} \cos \frac{2 \pi t}{T} dt \\
&=& A^2 \left( \sum_{n=1}^N h_{nm} \cos(\theta_n + \theta_{nm}) \right)^2 \frac{T}{2} +
\\
&& A^2 \left( \sum_{n=1}^N h_{nm} \sin(\theta_n + \theta_{nm}) \right)^2 \frac{T}{2} + \\
&& A^2 \sum_{n=1}^N h_{nm} \cos(\theta_n + \theta_{nm}) \sum_{n=1}^N h_{nm} \sin(\theta_n + \theta_{nm}) \cdot 0 \\
&=& \frac{A^2 T}{2} \beta_m(\bm{\theta})
\end{eqnarray*}
Thus, the total received power at all receivers is
\begin{eqnarray*}
E_{NM} = \sum_{m=1}^M E_{NM}(m) = \frac{A^2 T}{2} \sum_{m=1}^M \beta_m(\bm{\theta}) \; .
\end{eqnarray*}
Then we divide it by $\frac{A^2 h_{11}^2 T}{2}$ to obtain $G(\bm{\theta})$.

\section*{Appendix C -- Proof of Corollary~\ref{coro:upperbound}}

We obtain an upper bound for $G(\bm{\theta})$ as follows. We have
\begin{eqnarray*}
&& \hspace{-8mm} \beta_m(\bm{\theta}) \\
&\hspace{-5mm} \le& \hspace{-3mm} N \sum_{n=1}^N h_{nm}^2 \cos^2 (\theta_n + \theta_{nm}) +
N \sum_{n=1}^N h_{nm}^2 \sin^2 (\theta_n + \theta_{nm}) \\
&\hspace{-5mm} =& \hspace{-3mm} N \sum_{n=1}^N h_{nm}^2 (\cos^2 (\theta_n + \theta_{nm}) + \sin^2 (\theta_n + \theta_{nm})) \approx N^2 h_{11}^2 ,
\end{eqnarray*}
where the equality in $\le$ holds if
\begin{eqnarray*}
\cos⁡(\theta_1+ \theta_{1m})= \cos⁡(\theta_2+ \theta_{2m})= \cdots =\cos⁡(\theta_N+ \theta_{Nm}) \\
\sin⁡(\theta_1+ \theta_{1m})= \sin⁡(\theta_2+ \theta_{2m})= \cdots =\sin⁡(\theta_N+ \theta_{Nm}) \; ,
\end{eqnarray*}
and the approximation holds if
\begin{eqnarray*}
h_{nm}^2 \approx h_{11}^2 & m=1, \cdots, M \; .
\end{eqnarray*}
Thus, the gain at receiver $m$, $E_{NM}(m)/\frac{A^2 h_{11}^2 T}{2} = \frac{A^2 T}{2} \beta_m(\bm{\theta})/\frac{A^2 h_{11}^2 T}{2}$, can be up to $\frac{1}{h_{11}^2} N^2 h_{11}^2 = N^2$ times if phase coherence holds and channel gains are
similar, i.e.,
\begin{eqnarray*}
&& \theta_1+ \theta_{1m}= \cdots =\theta_N+ \theta_{Nm} \\
&& h_{nm}^2 \approx h_{11}^2 , \:\:\:  n=1, \cdots, N \; .
\end{eqnarray*}
Note that for coherent beamforming scenarios, the transmitters are close to each other and receivers are close to each other while the distance between transmitters and receivers is large.
For such a scenario, we can assume $h_{nm}^2 \approx h_{11}^2$.
If the above two conditions hold for all receivers, the total gain by all receivers can be up to $G_{\text{UB}} = N^2 M$ (such that $G(\bm{\theta}) \leq G_{\text{UB}}$ for any $\bm{\theta}$).

\section*{Appendix D -- Proof of Theorem~\ref{theorem:sirgain}}

The total received energy in one period is
\begin{eqnarray*}
E_{\mathcal{T}_k,\mathcal{R}_k} = \frac{A^2 T}{2} \rho_k(\bm{\theta}) \; .
\end{eqnarray*}
At the same time, transmitters for other data streams interfere at receivers for data stream $k$.
The interference in one period is given by
\begin{eqnarray*}
I_{\bm{\mathcal{T}},\mathcal{R}_k} &=& \sum_{m \in \mathcal{R}_k} \sum_{n \in \mathcal{T}_l}^{l \ne k} \frac{A^2 h_{nm}^2}{2} T
= \frac{A^2 T}{2} \sum_{m \in \mathcal{R}_k} \sum_{n \in \mathcal{T}_l}^{l \ne k} h_{nm}^2 \; ,
\end{eqnarray*}
where $\bm{\mathcal{T}}$ is the set of all $\mathcal{T}_k$.
Thus, the achieved SIR is
\begin{eqnarray*}
R_{\bm{\mathcal{T}},\mathcal{R}_k} =
\frac{\frac{A^2 T}{2} \rho_k(\bm{\theta})}{\frac{A^2 T}{2} \sum_{m \in \mathcal{R}_k} \sum_{n \in \mathcal{T}_l}^{l \ne k} h_{nm}^2}
= \frac{\rho_k(\bm{\theta})}{\sum_{m \in \mathcal{R}_k} \sum_{n \in \mathcal{T}_l}^{l \ne k} h_{nm}^2} .
\end{eqnarray*}
If coherent communication is not applied, the received energy in one period is given by $\frac{A^2 h_{s_k d_k}^2}{2} T$, where $h_{s_k d_k}^2$ is the channel gain from the source to the destination of data stream $k$.
The interference in one period is $\sum_{1 \le n \le K}^{n \ne k} \frac{A^2 h_{s_n d_k}^2}{2} T$.
Thus, the achieved SIR is
\begin{eqnarray*}
R_{\{ \{s_1\}, \cdots, \{s_K\} \}, \{d_k\}} &=& \frac{\frac{A^2 h_{s_k d_k}^2}{2} T}{\sum_{1 \le n \le K}^{n \ne k} \frac{A^2 h_{s_n d_k}^2}{2} T} \\
&=& \frac{h_{s_k d_k}^2}{\sum_{1 \le n \le K}^{n \ne k} h_{s_n d_k}^2} \; .
\end{eqnarray*}
Based on the above two SIR values, the coherent communication gain is
\begin{eqnarray*}
G_k(\bm{\mathcal{T}}, \bm{\mathcal{R}}, \bm{\theta}) &=& \frac{\rho_k(\bm{\theta})}{\sum_{m \in \mathcal{R}_k} \sum_{n \in \mathcal{T}_l}^{l \ne k} h_{nm}^2 \frac{h_{s_k d_k}^2}{\sum_{1 \le n \le K}^{n \ne k} h_{s_n d_k}^2}}
\nonumber \\
&=&
\frac{\sum_{1 \le n \le K}^{n \ne k} h_{s_n d_k}^2 \rho_k(\bm{\theta})}{h_{s_k d_k}^2 \sum_{m \in \mathcal{R}_k} \sum_{n \in \mathcal{T}_l}^{l \ne k} h_{nm}^2} \; .
\end{eqnarray*}


\begin{thebibliography}{99}

\bibitem{Shi18:coherent}
Y.~Shi, S.~Ponnaluri, Y.E.~Sagduyu, and E.v.~Doorn,
``Distributed coherent group communications,"
in \emph{Proc. IEEE Military Communications Conference (MILCOM),}
Los Angeles, CA, pp. 88--93, Oct.~29--31, 2018.

\bibitem{Bidigare13}
P.~Bidigare, D. R.~Brown, S.~Kraut, and U.~Madhow,
``MIMO channel prediction results on outdoor collected data,"
\emph{Asilomar Conference on Signals, Systems and Computers (ACSSC)}, Pacific Grove, CA, Nov. 2013.

\bibitem{Mudumbai09}
R.~Mudumbai, D. R.~Brown, U.~Madhow, and H.V.~Poor,
``Distributed transmit beamforming: Challenges and recent progress,"
\emph{IEEE Commun. Mag.},
vol.~47, no.~2, pp.~102--110, Feb. 2009.

\bibitem{Mudumbai10}
R.~Mudumbai, J.~Hespanha, U.~Madhow, and G.~Barriac,
``Distributed transmit beamforming using feedback control,"
\emph{IEEE Trans. Inf. Theory},
vol.~56, no.~1, pp.~411--426, Jan. 2010

\bibitem{Ochiai05}
H.~Ochiai, P.~Mitran, H.V.~Poor, and V.~Tarokh,
``Collaborative beamforming for distributed wireless ad hoc sensor networks,"
\emph{IEEE Trans. on Signal Process.},
vol.~53, no.~11, pp.~4110-4124, Oct. 2005.

\bibitem{Jayaprakasam17:sensor}
S.~Jayaprakasam, S.K.A.~Rahim, and C.Y.~Leow,
``Distributed and collaborative beamforming in wireless sensor networks: Classifications, trends, and research directions,"
\emph{IEEE Commun. Surveys Tuts.},
vol.~19, no.~4, pp.~2092--2116, June~2017.

\bibitem{Marques08:MISO}
A.G.~Marques, X.~Wang, and G.B.~Giannakis,
``Minimizing transmit power for coherent communications in wireless sensor networks with finite-rate feedback,"
\emph{IEEE Trans. on Signal Process.},
vol.~56, no.~9, pp.~4446--4457, Sep.~2008.

\bibitem{Nanzer17:MISO}
J.A.~Nanzer, R.L.~Schmid, T.M~ Comberiate, and J.E.~Hodkin,
``Open-loop coherent distributed arrays,"
\emph{IEEE Trans. Microw. Theory and Techn.},
vol.~65, no.~5, pp.~1662--1672, 2017.

\bibitem{Song09:MISO}
S.~Song, J.S.~Thompson, P.-J.~Chung, P.M.~Grant,
``Probability of error for BPSK modulation in distributed beamforming with phase errors,"
 \emph{ITG Workshop on Smart Antennas,}
Berlin, Germany, Feb.~16--18, 2009.

\bibitem{Zarifi10:MISO}
K.~Zarifi, A.~Ghrayeb, and S.~Affes,
``Distributed beamforming for wireless sensor networks with improved graph connectivity and energy efficiency,"
\emph{IEEE Trans. on Signal Process.},
vol.~58, no.~3, pp.~1904--1921, Mar.~2010.

\bibitem{Scherber13:MISO}
D.~Scherber, P.~Bidigare, R.~O'Donnell, M.~Rebholz, M.~Oyarzun, C.~Obranovich, W.~Kulp, D.~Chang, and D.R.~Brown III,
``Coherent distributed techniques for tactical radio networks: Enabling long range communications with reduced size, weight, power and cost,"
\emph{IEEE MILCOM,}
San Diego, CA, Nov.~18--20, 2013.

\bibitem{Quitin13:MISO}
F.~Quitin, M.M.U.~Rahman, R.~Mudumbai and U.~Madhow,
``A scalable architecture for distributed transmit beamforming with commodity radios: design and proof of concept,"
\emph{IEEE Trans. on Wireless Commun.},
vol.~12, no.~3, pp.~1418--1428, Mar.~2013.

\bibitem{Hu16:MISO}
P.-H.~Hu, P.-H.~Tseng, Y.-Y.~Guo, and C.-C.~Wei,
``Distributed transmit beamforming algorithms for unsynchronized OFDM systems with timing offset,"
\emph{IEEE Commun. Lett.},
vol.~20, no.~9, pp.~1788--1791, 2016.

\bibitem{Bidigare15}
T. P.~Bidigare, U.~Madhow, D. R.~Brown, R.~Mudumbai, A.~Kumar, B.~Peiffer, and S.~Dasgupta,
``Wideband distributed transmit beamforming using channel reciprocity and relative calibration,",
\emph{ACSSC},
Pacific Grove, CA, Nov. 2015.

\bibitem{Gencel1511:MISO}
M.F.~Gencel, M.E.~Rasekh, and U.~Madhow,
``Noise-resilient scaling for wideband distributed beamforming,"
\emph{ACSSC},
Pacific Grove, USA, Nov.~2015.

\bibitem{Gencel1506:MISO}
M.F.~Gencel, M.E.~Rasekh, and U.~Madhow,
``Scaling wideband distributed transmit beamforming via aggregate feedback,"
\emph{IEEE International Conference on Communications (ICC)},
London, UK, June~2015.

\bibitem{Deng16:radar}
H.~Deng, G.~Zhe, and B.~Himed,
``MIMO radar waveform design for transmit beamforming and orthogonality,"
\emph{IEEE Trans. Aerosp. and Electro. Syst.},
vol.~52, no.~3, pp.~1421--1433, 2016.



\bibitem{Shokouhifar17:clustering}
M.~Shokouhifar and A.~Jalali,
``Optimized sugeno fuzzy clustering algorithm for wireless sensor networks,"
\emph{Engineering Applications of Artificial Intelligence,}
2017.

\bibitem{Abuarqoub17:clustering}
A.~Abuarqoub, M.~Hammoudeh, B.~Adebisi, S.~Jabbar, A. Bounceur, and H.~Al-Bashar,
``Dynamic clustering and management of mobile wireless sensor networks,"
\emph{Elsevier Computer Networks,} vol.~117, no.~22, pp.~62-75,  Apr. 2017.

\bibitem{Cenedese17:clustering}
A.~Cenedese, M.~Luvisotto, and G.~Michieletto,
``Distributed clustering strategies in industrial wireless sensor networks,"
\emph{IEEE Trans. Ind. Informat.},
vol.~13, no.~1, pp.~228--237, Feb.~2017.

\bibitem{Rao17:clustering}
P.C.S.~Rao and H.~Banka,
``Novel chemical reaction optimization based unequal clustering and routing algorithms for wireless sensor networks,"
\emph{Springer Wireless Networks,}
vol.~23, no.~3, pp.~759-–778, April~2017.


\bibitem{Papadogiannis08:clustering}
A.~Papadogiannis, D.~Gesbert, and E.~Hardouin,
``A dynamic clustering approach in wireless networks with multi-cell cooperative processing,"
\emph{IEEE ICC},
Beijing, China, May~19--23, 2008.


\bibitem{Dai14:clustering}
B.~Dai and W.~Yu,
``Sparse beamforming and user-centric clustering for downlink cloud radio access network,"
\emph{IEEE Access},
vol.~2, pp.~1326--1339, Oct.~2014.

\bibitem{Ramamonjison14:clustering}
R.~Ramamonjison, A.~Haghnegahdar, and V.K.~Bhargava,
``Joint optimization of clustering and cooperative beamforming in green cognitive wireless networks,"
\emph{IEEE Trans. Wireless Commun.},
vol.~13, no.~2, pp.~982--997, Feb.~2014.

\bibitem{Zheng11:MISO-Clustering}
G.~Zheng, T.~Wei, and J.~Li,
``A MISO cooperative transmission algorithm for clustering wireless sensor network," \emph{International Conference on Electronics, Communications and Control,}
Ningbo, China,
Sep.~2011.


\bibitem{Pandula07}
S.~Pandula and B. P.~Paris,
``Performance analysis of MIMO receivers under imperfect CSIT,"
\emph{Conference on Information Systems and Sciences (CISS)}, Baltimore, MD, Mar.~14--16, 2007.

\bibitem{Pandula11}
S.~Pandula and B. P.~Paris,
``Distributed multi-cell power control in transmit beamforming systems using intercell-interference plus noise estimates,"
\emph{CISS},
Baltimore, MD, Mar.~23--25, 2011.

\bibitem{Alkhateeb14}
A.~Alkhateeb, O. E.~Ayach, G.~Leus, and R. W.~Heath,
``Channel estimation and hybrid precoding for millimeter wave cellular systems,"
\emph{IEEE J. Sel. Topics Signal Process.},
pp.~831--846, Oct. 2014.



\bibitem{Peiffer16:syn}
B.~Peiffer, R.~Mudumbai, S.~Goguri, A.~Kruger, and S.~Dasgupta,
``Experimental demonstration of retrodirective beamforming from a fully wireless distributed array,"
\emph{IEEE MILCOM,}
pp.~442--447, 2016.

\bibitem{Quitin12:syn}
F.~Quitin, U.~Madhow, M.M.U.~Rahman, and R.~Mudumbai,
``Demonstrating distributed transmit beamforming with software-defined radios,"
in \emph{Proc.~IEEE International Symposium on a World of Wireless, Mobile and Multimedia Networks (WoWMoM),}
3 pages, 2012.


\bibitem{Ponnaluri2017}
S. P. Ponnaluri, B. Azimi-Sadjadi, Y. E. Sagduyu, and W. Phoel, ``Distributed Interference Cancellation," \emph{IEEE Wireless Communications and Networking Conference (WCNC)}, San Fransisco, CA, Mar. 2017.

\bibitem{Ponnaluri15}
S. P.~Ponnaluri, S.~Soltani, Y.~Shi, and Y. E.~Sagduyu,
``Spectrum efficient communications with multiuser MIMO, multiuser detection and interference alignment,"
\emph{IEEE MILCOM,} Tampa, FL, Oct.~26--28, 2015.

\bibitem{Erpek18:MIMO}
T.~Erpek, Y. E.~Sagduyu, Y.~Shi, and S.~Ponnaluri,
``Rate optimization with distributed network coordination of multiuser MIMO communications," \emph{IEEE Vehicular Technology Conference},
Chicago, IL, Aug. 2018.

\bibitem{Erpek19}
T.~Erpek, Y. E.~Sagduyu, Y.~Shi, and S.~Ponnaluri,
``Network control and rate optimization for multiuser MIMO communications," \emph{Ad Hoc Networks Journal}, vol. 85, pp.~92--102, Mar. 2019.


\bibitem{Shi08}
Y.~Shi, Y.T.~Hou, and H.D.~Sherali,
``Cross-layer optimization for data rate utility problem in UWB-based ad hoc networks,"
\emph{IEEE Transactions on Mobile Computing},
vol.~7, issue~6, pp.~764--777, June~2008.

\bibitem{Shi09}
Y.~Shi, Y.T.~Hou, and H.~Zhou,
``Per-node based optimal power control for multi-hop cognitive radio networks,"
\emph{IEEE Transactions on Wireless Communications},
vol.~8, no.~10, pp.~5290--5299, Oct.~2009.

\bibitem{Lei12}
L. Yang, Y. E. Sagduyu, J. Zhang, and J. H. Li, ``Distributed stochastic power control in ad hoc networks: A nonconvex optimization case," \emph{EURASIP Journal on Wireless Communications and Networking}, 2012.

\bibitem{Chen05}
Q.~Chen, and Z.~Niu,
``A game-theoretical power and rate control for wireless ad hoc networks with step-up price,"
\emph{IEICE Transactions on Communications},
vol.~88, no.~9, pp.~3515--3523, 2005.

\bibitem{Heikkinen06}
T.~Heikkinen,
``A potential game approach to distributed power control and scheduling,"
\emph{Computer Networks},
vol.~50, no.~13, pp.~2295--2311, 2006.

\bibitem{ElNainay08}
M.Y.~El Nainay, D.H.~Friend, and A.B.~MacKenzie,
``Channel allocation \& power control for dynamic spectrum cognitive networks using a localized island genetic algorithm,"
\emph{Proc.~IEEE Symposium on New Frontiers in Dynamic Spectrum Access Networks},
Chicago, IL,
Oct.~2008.

\bibitem{Guo11}
S.~Guo, C.~Dang, and X.~Liao,
``Joint opportunistic power and rate allocation for wireless ad hoc networks: An adaptive particle swarm optimization approach,"
\emph{Journal of Network and Computer Applications}, vol.~34, no.~4 pp.~1353--1365, 2011.


\end{thebibliography}
\end{document}